\documentstyle[12pt]{article}

\begin{document}

\title{The ``Nernst Theorem'' and Black Hole Thermodynamics}
\author{Robert M. Wald\\
         {\it Enrico Fermi Institute and Department of Physics}\\
         {\it University of Chicago}\\
         {\it 5640 S. Ellis Avenue}\\
         {\it Chicago, Illinois 60637-1433}}
\maketitle

\begin{abstract}

The Nernst formulation of the third law of ordinary thermodynamics
(often referred to as the ``Nernst theorem'') asserts that the
entropy, $S$, of a system must go to zero (or a ``universal
constant'') as its temperature, $T$, goes to zero. This assertion is
commonly considered to be a fundamental law of thermodynamics.  As
such, it seems to spoil the otherwise perfect analogy between the
ordinary laws of thermodynamics and the laws of black hole mechanics,
since rotating black holes in general relativity do not satisfy the
analog of the ``Nernst theorem''. The main purpose of this paper is to
attempt to lay to rest the ``Nernst theorem'' as a law of
thermodynamics. We consider a boson (or fermion) ideal gas with its
total angular momentum, $J$, as an additional state parameter, and we
analyze the conditions on the single particle density of states,
$g(\epsilon,j)$, needed for the Nernst formulation of the third law to
hold. (Here, $\epsilon$ and $j$ denote the single particle energy and
angular momentum.) Although it is shown that the Nernst formulation of
the third law does indeed hold under a wide range of conditions, some
simple classes of examples of densities of states which violate the
``Nernst theorem'' are given. In particular, at zero temperature, a
boson (or fermion) gas confined to a circular string (whose energy is
proportional to its length) not only violates the ``Nernst theorem''
also but reproduces some other thermodynamic properties of an extremal
rotating black hole.

\end{abstract}
\newpage

\section{Introduction}

Nearly twenty five years ago a, remarkable relationship was
established \cite{bch} between the
ordinary laws of thermodynamics and certain laws of black hole
physics. This relationship was then greatly enhanced by the
discovery \cite{h2} that black holes radiate as perfect black bodies,
and by strong evidence for the validity of the ``generalized second
law'' \cite{b1}-\cite{fp}; see, e.g., \cite{w4}, \cite{w5} for
comprehensive reviews.

However, one apparent blemish has existed on this otherwise seemingly
perfect relationship. The Nernst formulation of the third law of
thermodynamics asserts that entropy, $S$, of a system must go to zero
(or a ``universal constant'') as its temperature, $T$, approaches
absolute zero. On the other hand, for Kerr black holes in general
relativity, the entropy is given by
\begin{equation}
S = A/4 = 2 \pi [M^2 +(M^4 - J^2)^{1/2}] ,
\label{A}
\end{equation}
and the temperature is given by
\begin{equation}
T = \kappa/2 \pi = 
\frac{(M^4 - J^2)^{1/2}}{4 \pi M [M^2 +(M^4 - J^2)^{1/2}]}
\label{T}
\end{equation}
where $M$ and $J$ denote, respectively, the mass and angular momentum
of the black hole. (Here and throughout this paper, we use units where
$G = c = \hbar = k = 1$.) Thus, absolute zero temperature corresponds
to the ``extremal limit''
\begin{equation}
J = M^2 .
\label{M0}
\end{equation}
The entropy at absolute zero temperature is thus
\begin{equation}
S = 2 \pi |J| ,
\label{S0}
\end{equation}
which is nonvanishing and, furthermore, has a functional dependence on
the state parameter $J$, so it does not approach a ``universal
constant''. Thus, the Kerr black holes stand in blatant violation of
the black hole mechanics analog of the ``Nernst theorem''.

This failure of the ``Nernst theorem'' to hold in black hole mechanics
has not generally been viewed with alarm by most researchers because
it is clear that the Nernst formulation of the third law does not have
the same fundamental status in thermodynamics as the first or second
laws (see, e.g., section 9.4 of the standard text of Huang
\cite{huang} for a clear statement of this view). Indeed, the Nernst
formulation of the third law does not hold at all in classical
physics, failing even for a classical ideal gas. In quantum
statistical physics, the ``Nernst theorem'' corresponds to a claim
about the behavior of the density of states, $n(E)$, as the total
energy of the system goes to its minimum possible value (or, more
precisely, as a statement about the extrapolation to minimum energy of
the higher energy, continuum approximation to $n(E)$; see
\cite{huang}). It is not difficult to concoct examples where $n(E)$ is
such that the Nernst formulation of the third law is violated. For
example, a system comprised by particles with spin but having no spin
interaction energy -- so that the ground state is highly degenerate --
will violate the ``Nernst theorem''.

Nevertheless, most such counterexamples to the Nernst formulation of
the third law seem rather contrived, and the fact that it has been
empirically found to hold for all systems studied in the laboratory
provides evidence that it might hold for all ``physically reasonable''
systems. If so, this would suggest that there might be something
``exotic'' about the the thermodynamic properties of extremal rotating
black holes.

In this paper we shall investigate this issue by studying the Nernst
formulation of the third law for a very non-exotic class of
thermodynamic systems: ideal boson gases. To keep the system as simple
as possible -- and, in particular, to avoid complications resulting
from Bose-Einstein condensation -- we shall assume that, as in the
case of the photon gas, particle number is not conserved;
equivalently, the chemical potential of the gas will be assumed to
vanish. However, we will assume that the gas is confined by an axially
symmetric box (or potential), so that its total angular momentum, $J$,
is conserved, and we will take $J$ and the total energy, $E$ to be the
state parameters of the system. The thermodynamic properties of the
gas are then determined by the single particle density of states,
$g(\epsilon,j)$, where $\epsilon$ and $j$ denote, respectively, the
single particle energy and angular momentum. In order to facilitate
our calculations, we shall further assume that $g(\epsilon,j)$ is
sufficiently ``non-exotic'' that the appropriate canonical ensemble --
modified to include angular momentum -- can be defined (at least at
low temperatures). This requires that $g(\epsilon,j)$ not grow more
rapidly than exponentially in $\epsilon$, and that the single particle
angular momentum to energy ratio be bounded, i.e., that $\Omega_{\pm}
> 0$, where
\begin{equation}
(\Omega_{\pm})^{-1} \equiv \sup (\pm j)/\epsilon .
\label{Ompm}
\end{equation}
Thus, we have $g(\epsilon,j) = 0$ unless
\begin{equation}
- \epsilon/\Omega_- \leq j \leq \epsilon/\Omega_+ .
\label{je}
\end{equation}
(Note that this condition holds for a system of free particles
confined to within a (cylindrical) radius $R$ of the symmetry axis,
with $\Omega_{\pm} = 1/R$.) We then pose the following two questions:
(i) What properties of $g(\epsilon,j)$ are required in order that the
Nernst formulation of the third law be violated, i.e., so that
$S(T,J)$ approaches a non-zero limit (which depends upon $J$) as $T
\rightarrow 0$? (ii) Can these conditions be achieved for any classes
of ``physically reasonable'' ideal gas systems?

Of course, even if the answer to (ii) were ``no'', this would not mean
that extremal Kerr black holes necessarily display any ``unphysical''
or ``exotic'' thermodynamic behavior, since there is no reason to
expect that their behavior could be properly modeled by an ideal boson
gas. Indeed, with the restrictions placed on the density of states
needed to define the ordinary canonical ensemble, it is impossible to
get negative heat capacities, as occurs for black holes with
sufficiently small angular momentum. There is nothing ``unphysical''
or ``exotic'' about systems with negative heat capacities; for
example, ordinary self-gravitating stars in Newtonian gravity have
negative heat capacities. However, the simple ideal gas systems we
consider here are not adequate to model this behavior. There is no
reason, a priori, to believe that they should be adequate to model the
violations of ``Nernst's theorem'' displayed by extremal Kerr black
holes. Nevertheless, it is of interest to see how close one can come
to modeling the thermodynamic behavior of extremal Kerr black holes
with ideal boson gas systems.

As we shall see in the next section, for a violation of ``Nernst's
theorem'', it is sufficient (and, as explained there, ``nearly
necessary'') that there exist single particle states which achieve the
limit (\ref{je}), i.e., that (for positive $J$) there exist states
which satisfy $\epsilon = \Omega_+ j$ exactly. No such states exist
for a free boson gas confined by a spherical box in two or higher
spatial dimensions, and such systems satisfy the Nernst formulation of
the third law even when they are rotating. (We will explicitly
calculate the low temperature behavior of a rotating gas in the next
section.) However, massless ideal gases in one spatial dimension and
ideal gases in ``zero spatial dimensions'' (i.e., spin systems) do
have states for which $\epsilon = \Omega_+ j$, and they violate the
``Nernst theorem'' when angular momentum is taken into account. Thus,
violations of the ``Nernst theorem'' -- which are qualitatively very
similar the violations of the ``Nernst theorem'' for Kerr black holes
-- do occur for some simple systems comprised by ideal gases with
angular momentum, although the one (or zero) dimensionality of such
systems seems essential.

Encouraged by this result, we may ask if the detailed thermodynamic
properties of extremal Kerr black holes given by eqs.(\ref{M0}) and
(\ref{S0}) also can be modeled by ideal gas systems. As we shall see
in the next section, for $J > 0$ the ideal gas systems will
automatically satisfy $E = \Omega_+ J$ at zero temperature, rather
than $E \propto J^{1/2}$, as in eq.(\ref{M0}). However, if we modify
the model of a one-dimensional boson gas confined to a ring of radius
$R$ by simply treating $R$ itself as an additional classical dynamical
variable, and if we also attribute an additional energy proportional
to $R$ (due to ``string tension'') to the total energy $E$, then the
behavior $E \propto J^{1/2}$ is obtained -- in agreement with
(\ref{M0}). However, the behavior $S \propto J$ at zero temperature
(see eq.(\ref{S0})) seems much more difficult to model, as it appears
to require the density of states, $n(j)$, at $\epsilon = \Omega_+ j$
to grow exponentially with $j$. (A collection of massless boson gases
would have a constant $n(j)$, which leads to the behavior $S \propto
J^{1/2}$ at zero temperature.) Nevertheless, it seems remarkable that
such a simple model can come so close to mimicking the thermodynamic
behavior of extremal Kerr black holes.

This investigation was stimulated by the recent success in modeling
the thermodynamic behavior of certain extremal charged black holes
(namely, those which saturate the ``BPS bound'') in string theory
\cite{sv}. These results already provide a counterexample to the
``Nernst theorem'' for a particular system in the class considered
here, since the degrees of freedom which contribute to the entropy in
the weak coupling string model correspond to that of a free,
one-dimensional gas. In the present investigation, we consider
general ideal boson gas systems -- not restricted by any models
arising from string theory.\footnote{The philosophy of the present
paper bears some similarity with the philosophy adopted in a recent
paper of Maldecena and Strominger \cite{ms}, who study the emission
properties of nearly BPS, slowly rotating black holes and deduce from
those properties some aspects of the effective string theory
description of such black holes. However, there does not appear to be
any overlap in the contents of that paper and the present paper.} The
one (or zero) dimensionality of the models we find which violate the
``Nernst theorem'' is a conclusion, rather than an input, of our
analysis.

Finally, we note that, for definiteness, we shall consider an ideal
boson gas at zero chemical potential in our analysis. However, the
analysis of an ideal fermion gas (at zero chemical potential) would
proceed in complete parallel -- with merely some sign changes in
various expressions -- and the conclusions in the fermion case would
be unaltered.

\section{The thermodynamical properties of a rotating boson gas at 
low temperatures}

Consider an ideal boson gas, confined by a potential (or ``box'')
which is axially symmetric. Then the angular momentum about the
symmetry axis is conserved, and the single-particle states of the gas
can be labeled by their energy, $\epsilon$, and angular momentum, $j$,
about the symmetry axis. We shall assume that the single particle
Hamiltonian is positive, and that the minimum energy single particle
state is $\epsilon_0 > 0$. (This ensures that the ``vacuum state'' is
the unique ground state of the system. If there existed any single
particle states with $\epsilon = 0$, the ground state of system would
be highly degenerate and the Nernst formulation of the third law would
be trivially violated even when the total angular momentum vanishes.)
Let $G(\epsilon,j)$ denote the number of states with energy $\leq
\epsilon$ and angular momentum $\leq j$. Thus, $G$ is non-negative, is
a monotone increasing function of $\epsilon$ and $j$, and satisfies
$G(0,j) = 0$. The density of states, $g(\epsilon,j)$, is defined by
\begin{equation}
g(\epsilon,j) = \frac{\partial^2 G}{\partial \epsilon \partial j} .
\label{dos}
\end{equation}
In reality, on account of the discreteness of states, $G(\epsilon,j)$
is a piecewise constant function and, correspondingly, $g$ is a sum of
delta-functions, but (following standard practice) in our expressions
we will treat both of them as ``continuum'' (though {\em not}
necessarily continuous) variables, i.e., we will write down integral
expressions rather than sums in our formulas below. However, all of
our formulas will continue to make sense if $g$ is taken to be a sum
of delta-functions (or has delta-function contributions in addition to
contributions which are treated as being continuous).

We will assume that, as for the case of a photon gas, particle number
in our boson gas is not conserved, i.e., that particles can be created
freely, at no ``cost'' other than the energy and angular momentum
required to create them. (This corresponds to a vanishing chemical
potential of the gas.)  Thus, the state variables will not include the
number of particles and will be taken to be simply $E$ and $J$. Given
only that $G(\epsilon,j)$ is bounded in $j$ at each $\epsilon$ (i.e.,
that for each $\epsilon$ there are only a finite number of single
particle states with energy $<\epsilon$), the microcanonical ensemble
appropriate to fixing the total energy, $E$, and total angular
momentum, $J$, is well defined. The entropy, $S(E,J)$, may then be
defined as $S(E,J) = \ln N(E,J)$, where $N(E,J)$ denotes the number of
states of the total system ({\em not} single particle states) with
total energy between $E$ and $E + \Delta E$ and total angular momentum
between $J$ and $J + \Delta J$. However, use of the microcanical
ensemble is not very convenient for most calculations, and the entropy
of systems is usually computed in the context of the canonical
ensemble.

To obtain the appropriate canonical ensemble in the present case, we
proceed in close parallel to the derivation of the grand canonical
ensemble. We imagine that our system is able to exchange energy and
angular momentum with a ``heat bath/angular momentum reservoir''
(rather than a ``heat bath/particle reservoir'') characterized by
temperature $T = 1/\beta$ and angular velocity $\Omega$. (Here $T$ and
$\Omega$ are defined by their appearance in the first law of
thermodynamics for the reservoir, namely $dE = T dS + \Omega dJ$.) In
order that our ideal gas system be able to ``come to equilibrium''
with the reservoir (so that the canonical ensemble can be defined) it
is necessary to impose two additional restrictions on $G(\epsilon,j)$:
First, in the usual manner, we must have $G(\epsilon,j) \leq C
\exp(\alpha \epsilon)$ for some constants $C$ and $\alpha$, since
otherwise the system could indefinitely soak up energy from the
reservoir. Second, we must have $\Omega_+ > 0$ and $\Omega_- > 0$
(where $\Omega_+$ and $\Omega_-$ were defined by eq.(\ref{Ompm})
above), since otherwise the system could indefinitely soak up angular
momentum from the reservoir. In the following, we shall assume that
these conditions are satisfied -- so that the canonical ensemble is
well defined for $T < 1/\alpha$ and $-\Omega_- < \Omega <
\Omega_+$. We then shall use canonical ensemble methods to compute
$S(T,J)$. As usual, the canonical ensemble is equivalent to the
microcanonical ensemble for the purposes of computing the entropy and
other thermodynamic quantities for the system provided that the energy
and angular momentum fluctuations in the canonical ensemble are
sufficiently small.\footnote{At extremely low temperatures, the
microcanonical and canonical ensembles need not be
equivalent. However, as emphasized in \cite{huang}, the Nernst
formulation of the third law really refers to the extrapolation to $T
= 0$ of the formula for the entropy which applies at temperatures
which are sufficiently high that the two ensembles should be
equivalent. Thus, it is appropriate to use the canonical ensemble for
our calculations here even if the two ensembles are not equivalent at
$T = 0$.}

In exact parallel with the grand canonical ensemble, in our ``angular
momentum modified canonical ensemble'', all thermodynamic quantities
can be derived in a straightforward manner from a partition function
$Z(\beta,\Omega)$. For an ideal boson gas, $Z$ is given by,
\begin{equation}
\ln Z = - \int d\epsilon dj g(\epsilon,j) \ln[1 - \exp(-\beta[\epsilon
- \Omega j])] .
\label{Z}
\end{equation}
The (expected) angular momentum, $J$, is then given by
\begin{eqnarray}
J & = & \frac{1}{\beta} \frac{\partial \ln Z}{\partial \Omega}
\nonumber\\ 
& = & \int d\epsilon dj g(\epsilon,j) 
\frac{j}{\exp(\beta[\epsilon - \Omega j]) - 1} .
\label{J}
\end{eqnarray}
The (expected) energy, $E$, is determined by
\begin{eqnarray}
E - \Omega J & = & - \frac{\partial \ln Z}{\partial \beta}  \nonumber\\
& = & \int d\epsilon dj g(\epsilon,j) 
\frac{\epsilon - \Omega j}{\exp(\beta[\epsilon - \Omega j]) - 1} .
\label{E}
\end{eqnarray}
Finally, the entropy, $S$, is given by
\begin{eqnarray}
S & = & \ln Z + \beta (E - \Omega J)  \nonumber\\
& = & \ln Z - \beta  \frac{\partial \ln Z}{\partial \beta}. 
\label{S}
\end{eqnarray}
Equation (\ref{S}) yields the entropy as a function of $\beta$ and
$\Omega$. To obtain $S(\beta,J)$, we must solve eq.(\ref{J}) to
express $\Omega$ as a function of $\beta$ and $J$. Our task is to find
conditions on the density of states, $g(\epsilon,j)$, so that
$S(\beta,J)$ does not approach zero (or a ``universal constant'') when
$\beta \rightarrow \infty$ at fixed $J$.

In the following, we shall restrict attention to analyzing the case
where $J > 0$. (In particular, the case $J = 0$ will be excluded from
our analysis.) The states with $j$ near its maximal value
$\epsilon/\Omega_+$ will then play an important role in the behavior
of the gas as $\beta \rightarrow \infty$, and it useful to replace the
variable $\epsilon$ with the variable
\begin{equation}
y \equiv \epsilon - \Omega_+ j .
\label{y}
\end{equation}
The allowed ranges of $y$ and $j$ corresponding to the restrictions
(\ref{je}) are then
\begin{equation}
y \geq 0; \;  j \geq - y/(\Omega_+ +\Omega_-).
\label{yj}
\end{equation}
In addition, the condition $\epsilon \geq \epsilon_0$ yields
\begin{equation}
j \geq (\epsilon_0- y)/\Omega_+ .
\label{yj'}
\end{equation}
We define $H(y,j)$ to be the total number of states labeled by
$(y',j')$, such that $y' \leq y$ and $j' \leq j$. We define the
corresponding density of states, $h(y,j)$, by
\begin{equation}
h(y,j) = \frac{\partial^2 H}{\partial y \partial j} .
\label{h}
\end{equation}
Then, we have $h(\epsilon - \Omega_+ j, j) = g(\epsilon,j)$, although
the relationship between $H$ and $G$ is not quite as straightforward,
since the state counting in the two cases is being done over different
regions of single particle state space. In terms of our new variables,
the above formula (\ref{Z}) for $\ln Z$ becomes
\begin{eqnarray}
\ln Z & = & - \int dy dj h(y,j) 
\ln[1 - \exp(-\beta y - \beta[\Omega_+ - \Omega] j)]  \nonumber\\
& = & \sum_{n=1}^\infty \frac{1}{n} \int dy dj 
\frac{\partial^2 H}{\partial y \partial j} e^{-n \beta y} e^{- n \sigma j} ,
\label{Z2}
\end{eqnarray}
where, in the second line, we have made use of the series expansion
\begin{equation}
\ln[1- e^{-x}] = - \sum_{n=1}^\infty \frac{1}{n} e^{-nx} 
\label{series}
\end{equation}
and we have written
\begin{equation}
\sigma \equiv \beta (\Omega_+ - \Omega) .
\label{sigma}
\end{equation}
(Note that $\sigma > 0$ in order for the canonical ensemble to be
defined.)  The corresponding series expanded formulas for $J$ and $S$
in our new variables are
\begin{equation}
J = \sum_{n=1}^\infty \int dy dj  
\frac{\partial^2 H}{\partial y \partial j} j e^{-n \beta y} e^{- n \sigma j}
\label{J2}
\end{equation}
and
\begin{equation}
S = \sigma J + \sum_{n=1}^\infty \frac{1}{n} \int dy dj  
\frac{\partial^2 H}{\partial y \partial j} e^{-n \beta y} e^{- n\sigma j} 
+ \sum_{n=1}^\infty \beta \int dy dj  
\frac{\partial^2 H}{\partial y \partial j} y e^{-n \beta y} e^{- n \sigma j} 
\label{S2}
\end{equation}

We now integrate eqs.(\ref{J2}) and (\ref{S2}) by parts with respect
to both $y$ and $j$ (taking the ranges of both of these integrals to
be $-\infty$ to $\infty$). When we do so, no boundary terms arise from
the upper limits on account of the exponentially decaying terms $e^{-n
\beta y}$ and $e^{- n \sigma j}$, and no boundary terms arise from the
lower limits on account of the vanishing of $H(y,j)$ outside of the
range defined by eq.(\ref{yj}). We obtain
\begin{equation}
J = \sum_{n=1}^\infty n^2 \beta \sigma\int dy dj  
H(y,j) (j - \frac{1}{n\sigma}) e^{-n \beta y} e^{- n \sigma j} 
\label{J3}
\end{equation}
and
\begin{equation}
S =  \sigma J + \sum_{n=1}^\infty n^2 \beta^2 \sigma\int dy dj  
H(y,j) y e^{-n \beta y} e^{- n \sigma j} .
\label{S3}
\end{equation}
Finally, we introduce the new variables
\begin{equation}
w = n \sigma j, \;  z = n \beta y
\label{wz}
\end{equation}
to convert these expressions to the form
\begin{equation}
J = \frac{1}{\sigma} \sum_{n=1}^\infty \frac{1}{n} \int_0^\infty dz 
\int_{- \frac{z \sigma}{\beta(\Omega_- + \Omega_+)}}^\infty dw  
H(\frac{z}{n\beta},\frac{w}{n\sigma}) (w-1) e^{-z} e^{- w} 
\label{J4}
\end{equation}
and
\begin{equation}
S = \sigma J + \sum_{n=1}^\infty \frac{1}{n} \int_0^\infty dz 
\int_{- \frac{z \sigma}{\beta(\Omega_- + \Omega_+)}}^\infty dw  
H(\frac{z}{n\beta},\frac{w}{n\sigma}) z e^{-z} e^{- w} ,
\label{S4}
\end{equation}
where we have now explicitly inserted lower limits on the integrals to
remind the reader that $H$ vanishes outside the range defined by
eq.(\ref{yj}).  Note that since the second term on the right side of
eq.(\ref{S4}) is non-negative, we have
\begin{equation}
S \geq \sigma J .
\label{S5}
\end{equation}

We now show that for any fixed $J > 0$, $\sigma$ must remain bounded
from above when $\beta \rightarrow \infty$, i.e., $\Omega$ must
approach $\Omega_+$ at least as rapidly as $1/\beta$. Equivalently, we
have $\sigma_0 < \infty$ where
\begin{equation}
\sigma_0 \equiv \limsup_{\beta \rightarrow \infty} \sigma .
\label{sigma0}
\end{equation}
To see this, we note that by eq.(\ref{J4}) we have
\begin{equation}
J  \leq  \frac{1}{\sigma} \sum_{n=1}^\infty \frac{1}{n} \int_0^\infty dz 
\int_0^\infty dw  H(\frac{z}{n\beta},\frac{w}{n\sigma}) (w-1) e^{-z} e^{- w}
\label{J5}
\end{equation}
If there were a sequence $\beta_i \rightarrow \infty$ such that
$\sigma_i \rightarrow \infty$, then -- on account of the factor of
$1/\sigma$ together with the fact that $H$ is a monotone increasing
function of both of its arguments (and, hence, is monotone decreasing
along this sequence) -- the right side of eq.(\ref{J5}) would converge
to zero, in contradiction with the fact that $J > 0$.

A crucial factor in the behavior of $S$ at $T=0$ is whether or not
$\sigma_0 = 0$. If $\sigma_0 > 0$, then by eq.(\ref{S5}) we have
\begin{equation}
\limsup_{\beta \rightarrow \infty} S \geq \sigma_0 J > 0 ,
\label{S6}
\end{equation}
and the Nernst formulation of the third law will fail. On the other
hand, suppose that $\sigma_0 = 0$ (so that $\sigma \rightarrow 0$ as
$\beta \rightarrow \infty$, i.e., $\Omega$ approaches $\Omega_+$ more
rapidly than $1/\beta$). Then $\sigma J$ converges to zero, so we only
need worry about the second term on the right side of
eq.(\ref{S4}). However, in order to keep the right side of
eq.(\ref{J4}) from diverging as $\beta \rightarrow \infty$, it is
necessary that $H(\frac{z}{\beta},\frac{w}{\sigma})$ converge
pointwise to zero for all $z,w \geq 0$. (If not, then using the
monotonicity and positivity of $H$, the integrals on the right side of
eq.(\ref{J4}) would remain finite, but the $1/\sigma$ factor would
diverge.) If we knew, in addition, that for all $\beta$ we had
$H(\frac{z}{\beta},\frac{w}{\sigma}) \leq F(z,w)$ where $F$ is such
that $\int dz dw F(z,w) z e^{-z} e^{- w}$ converges, then we could use
the dominated convergence theorem to conclude that $S \rightarrow 0$
as $\beta \rightarrow \infty$. I have not attempted to give a complete
analysis of the conditions on $H$ which are necessary and sufficient
for the Nernst behavior to occur when $\sigma_0 = 0$, but it seems
clear that this ``normally'' will be the case (and possibly always is
the case, since I do not know of any counterexamples to the Nernst
behavior when $\sigma_0 = 0$).

What conditions on $H$ are necessary and sufficient to ensure that
$\sigma_0 > 0$, so that the Nernst formulation of the third law will
be violated? A sufficient condition is that $H(0,j) > 0$ for some $j$,
i.e., that there exists at least one single particle state which
actually achieves the limiting angular momentum $j = \epsilon
/\Omega_+$. To see this, we note that if we assume that $H(0,j) > 0$
for some $j$ but that $\sigma_0 = 0$, it follows immediately that
$H(\frac{z}{\beta},\frac{w}{\sigma})$ cannot converge pointwise to
zero. However, as in the arguments of the previous paragraph, this
yields a contradiction, since it implies that $J \rightarrow \infty$
as $\beta \rightarrow \infty$.

On the other hand, for a wide class of $H$'s, the condition that
$H(0,j) > 0$ for some $j$ also is necessary to have $\sigma_0 > 0$. In
particular, suppose that $H(y,j)$ is polynomially bounded in $j$ at
each $y$ in such a way that for $j \geq 0$ we have
\begin{equation}
H(y,j) \leq F(y) (1 + j^k)
\label{Hbnd}
\end{equation}
where $F(y)$ is continuous, is exponentially bounded at large $y$ (so
that the canonical ensemble is well defined at large $\beta$), and
satisfies $F(0) = 0$. This behavior encompasses a very wide class of
$H$'s such that $H(0,j) = 0$ for all $j$. Since $H$ is a monotone
increasing function of $y$, we may assume, without loss of generality,
that $F$ also is a monotone increasing function. By eq.(\ref{J5}), we
have
\begin{eqnarray}
J & \leq & \frac{1}{\sigma} \sum_{n=1}^\infty \frac{1}{n}
\int_0^\infty dz \int_0^\infty dw  F(\frac{z}{n\beta}) 
[1 + (\frac{w}{n\sigma})^k] (w -1) e^{-z} e^{- w}   \nonumber\\
& \leq & \frac{1}{\sigma^{(k+1)}} \Gamma(k+2) 
\sum_{n=1}^\infty \frac{1}{n^{(k+1)}}  
\int_0^\infty dz F(\frac{z}{n\beta}) e^{-z}  \nonumber\\
& \leq & \frac{C}{\sigma^{(k+1)}} 
\int_0^\infty dz F(\frac{z}{\beta}) e^{-z} 
\label{Jbnd}
\end{eqnarray}
where the monotone property of $F$ was used in the last line to obtain
$F(\frac{z}{n\beta}) \leq F(\frac{z}{\beta})$. However, as $\beta
\rightarrow \infty$, the functions $f_\beta(z) \equiv
F(\frac{z}{\beta})$ converge pointwise to $0$ and are ``dominated'' by
$F(z)$, so, by the dominated convergence theorem, the integral on the
right side of eq.(\ref{Jbnd}) converges to 0. Consequently, we must
have $\sigma_ 0 = 0$ in this case, as we desired to show.

If $H(y,j)$ is not polynomially bounded in $j$, then it is possible to
have $\sigma_0 > 0$ even if $H(0,j) = 0$ for all $j$. Indeed, if
$H(y,j) = F(y) e^{\lambda j}$ where $\lambda > 0$ and $F$ is as in the
previous paragraph, then it is not difficult to see from eq.(\ref{J4})
that $\sigma_0 = \lambda > 0$. However, I am not aware of any
circumstances under which $\sigma_0 > 0$ when $H(0,j) = 0$ for all $j$
and $H(y,j)$ is such that at fixed $y$, $H(y,j) e^{-\alpha j}$ is
bounded in $j$ for all $\alpha > 0$.

We now summarize our results. We have considered ideal boson gases
whose single particle states satisfy the restriction (\ref{je}). We
have shown above that {\em if there exist any single particle states
which actually achieve the maximal ratio of angular momentum to energy
-- namely $j/\epsilon = 1/\Omega_+$ -- then the Nernst formulation of
the third law will fail for $J > 0$}. In a limited class of other
circumstances -- in particular, when $H(y,j)$ grows exponentially with
$j$ -- the Nernst formulation of the third law also may fail even if
no single particle states satisfy $j/\epsilon = 1/\Omega_+$. However,
it appears that in the ``vast majority of cases'' -- and conceivably
all cases where $H(y,j) e^{-\alpha j}$ bounded in $j$ for all $\alpha
> 0$ -- the Nernst formulation of the third law holds when no single
particle states satisfy $j/\epsilon = 1/\Omega_+$.

A few simple examples are useful to illustrate these general results
and to gain insight into the conditions under which there are
states with $j/\epsilon = 1/\Omega_+$, so that the Nernst formulation
of the third law is violated. As a first example, consider a gas of
particles of a free, massless, scalar field in three dimensions,
confined by a spherical box of radius $R$, with Dirichlet boundary
conditions on the walls of the box. The spatial mode functions for the
particles are then of the form
\begin{equation}
\phi_{nlm} = j_l(k_{ln} r) Y_{lm}(\theta, \varphi)
\label{modes}
\end{equation}
where $k_{ln}$ is the $n$th value of $k$ such that $j_l(kR) = 0$. The
energy of the mode $\phi_{nlm}$ is $k_{ln}$ and its ($z$-)angular
momentum is $m$. (Recall that we are using units in which $\hbar =
1$.) Since $k_{ln} > (l + 1/2)/R$ (see, e.g., \cite{watson}), we have
$j/\epsilon < 1/R$ for all single particle states. However, since the
first zero, $k_{l1}$, satisfies \cite{watson}
\begin{equation}
\lim_{l \rightarrow \infty} \frac{k_{l1}}{l} = 1
\label{k1}
\end{equation}
we see that $\Omega_+ = 1/R$, and no single particle state actually
achieves the maximal angular momentum to energy ratio.

By the above arguments, the Nernst formulation of the third law should
hold in this example. To see this explicitly, we note that for large
$l$, the density of zeros of $j_l (x)$ is given by
\begin{equation}
\rho = \frac{1}{\pi} (1 - \frac{(l+1/2)^2}{x^2})^{1/2} .
\label{rho}
\end{equation}
(This result can be derived from formulas given in section 15.81 of
\cite{watson}.) Each $l$ contributes one state of ($z$-)angular
momentum $j$ (for integer $j$) if $l \geq |j|$ and zero states
otherwise. Hence, the density of states, $g(\epsilon,j)$, is given by
\begin{eqnarray}
g(\epsilon,j) & = & \frac{R}{\pi} \int_{|j|}^{\epsilon R - 1/2} 
(1 - \frac{(l+1/2)^2}{(\epsilon R)^2})^{1/2} dl \nonumber\\ 
& = & \frac{R^2 \epsilon}{2 \pi} \{\arccos(|j|/R\epsilon) - 
(|j|/R\epsilon)[1 - (j/R\epsilon)^2]^{1/2}\}
\label{g1}
\end{eqnarray}
In terms of the variable $y = \epsilon - j/R$, the density of states
$h(y,j)$ is
\begin{equation}
h(y,j) = \frac{R}{2 \pi} (j + Ry) 
\{\arccos(\frac{|j|}{j+Ry}) - \frac{|j|}{(j+Ry)^2}[2Rjy + R^2y^2]^{1/2}\} .
\label{h1}
\end{equation}
Taking into account the restriction (\ref{yj}), we see that
\begin{equation}
h(y,j) \leq C (|j| +Ry) \leq C' (1 + y) (1 + |j|) ,
\label{h1'}
\end{equation}
from which it follows immediately that
\begin{equation}
H(y,j) \leq C'' (y + y^2) (1 + j^2) ,
\label{H1'}
\end{equation}
which is of the form (\ref{Hbnd}). Thus, we have $\sigma_0 = 0$ in
this case.

The explicit behavior of the entropy of the rotating gas at low
temperatures can be calculated as follows. From eq.(\ref{h1}), we see
that for small $y$, we have
\begin{equation}
h(y,j) \approx \frac{2 \sqrt{2}}{3 \pi} R^{5/2} j^{-1/2} y^{3/2} .
\label{h2}
\end{equation}
Substituting this into eq.(\ref{J2}), we find that for $J > 0$ and
large $\beta$
\begin{eqnarray}
J & \approx & \frac{2 \sqrt{2}}{3 \pi} R^{5/2}\sum_{n=1}^\infty 
\int dy dj   j^{1/2} y^{3/2} e^{-n \beta y} e^{- n \sigma j}  \nonumber\\
& \approx & \frac{2 \sqrt{2}}{3 \pi} R^{5/2} \Gamma(5/2) 
\Gamma(3/2) \zeta(4) \beta^{-5/2} \sigma^{-3/2} \nonumber\\
& = & \frac{\sqrt{2} \pi^4}{360} R^{5/2} \beta^{-5/2} \sigma^{-3/2} .
\label{J7}
\end{eqnarray}
(Here $\zeta$ denotes the Riemann zeta function, and we have used the
values $\zeta(4) = \pi^4/90$, $\Gamma(3/2) = \sqrt{\pi}/2$, and
$\Gamma(5/2) = 3 \sqrt{\pi}/4$.) Thus, at large $\beta$, we have
\begin{equation}
\sigma \approx  \{\frac{\sqrt{2} \pi^4}{360}\}^{2/3} 
\frac{R^{5/3}}{\beta^{5/3} J^{2/3}} .
\label{sigma2}
\end{equation}
Substituting this into eq.(\ref{S2}), we find that as $T \rightarrow
0$ at fixed $J > 0$, we have
\begin{equation}
S \propto R^{5/3} J^{1/3} T^{5/3} \rightarrow 0 .
\label{S7}
\end{equation}
Thus, the Nernst formulation of the third law does indeed hold,
although $S$ goes to zero more slowly than in the case where the
angular momentum of the gas is not constrained (in which case $S
\propto R^3 T^3$ at all temperatures).

I have not succeeded in finding any simple examples of systems
violating the Nernst formulation of the third law which -- like the
case of a free boson gas in a spherical box -- satisfy the properties
that (i) the angular momentum carried by the particles is primarily
``orbital'' (as opposed to ``spin'') in character, and (ii) the
particles are not constrained to move exclusively in the
$\varphi$-direction. However, it is easy to find simple examples of
``zero-dimensional systems'' (i.e., spin systems) and one-dimensional
systems which violate the Nernst formulation of the third law.

As a simple example of a spin system which violates the Nernst
formulation of the third law, suppose that we have bosonic particles
of mass M and spin s, which can be located on any one of N ``lattice
sites''. (Again, the total number of such particles is taken to be
unconstrained.) Then the maximal angular momentum to energy ratio for
single particle states is $s/M$ (i.e., $\Omega_+ = M/s$), which is
attained by particles whose spin is aligned along the $z$-axis. In
this case, we clearly have $H(0,j) = 0$ for $j < s$, whereas $H(0,j) =
N$ for $j \geq s$. The states with $y = 0$ (i.e., $j = s$) will dominate
the low temperature behavior of the gas when $J = 0$. Thus, taking the
limit as $\beta \rightarrow \infty$ in eqs.(\ref{J4}) and (\ref{S4})
and performing the $z$-integrals, we find that at $T = 0$
\begin{equation}
J = \frac{1}{\sigma} \sum_{n=1}^\infty \frac{1}{n} \int_0^\infty dw  
H(0,\frac{w}{n\sigma}) (w-1) e^{- w} 
\label{J4'}
\end{equation}
and
\begin{equation}
S = \sigma J + \sum_{n=1}^\infty \frac{1}{n} \int_0^\infty dw  
H(0,\frac{w}{n\sigma}) e^{- w} .
\label{S4'}
\end{equation}
Consequently, in the present case, we have
\begin{eqnarray}
J & = & N \frac{1}{\sigma} \sum_{n=1}^\infty \frac{1}{n} 
\int_{sn\sigma}^\infty dw  (w-1) e^{- w}   \nonumber\\
& = & Ns \sum_{n=1}^\infty e^{-sn\sigma} \nonumber\\
& = & Ns \frac{1}{e^{s \sigma} - 1} .
\label{J5'}
\end{eqnarray}
Similarly, we get
\begin{equation}
S = \sigma J - N \ln(1-e^{-s \sigma}) \; .
\label{S5'}
\end{equation}
Eliminating $\sigma$, we find that at $T = 0$, we have
\begin{equation}
S = \frac{J}{s} \ln[1+ \frac{Ns}{J}] + N\ln[1+ \frac{J}{Ns}] ,
\label{S6'}
\end{equation}
which violates the Nernst formulation of the third law. Note that a
similar behavior of the entropy at $T = 0$ also should hold for any
system in which the angular momentum of the system is carried in
discrete ``vortex structures'', such as occurs in superfluid
helium. (Here, $N$ should correspond roughly to the number of vortex
structures that could occur in the superfluid helium without
overlapping. Presumably, we would need $J/s << N$ in order to have the
vortex structures present.) Thus, if the vortex structures in
superfluid helium persist to absolute zero temperature and can be
treated as non-interacting, that system should violate the Nernst
formulation of the third law. However, the entropy contributed by the
vortex structures should be negligible at temperatures achievable in
the laboratory.

Another simple example of a system which violates the Nernst
formulation of the third law is provided by a free, massless, gas of
scalar particles, which is confined to a one-dimensional ring of
radius $R$. The states in this case decompose into ``right movers''
and ``left movers'', and the density of states is simply
\begin{equation}
g(\epsilon,j) = \delta(\epsilon - j/R) + \delta(\epsilon + j/R) .
\label{g1'}
\end{equation}
Thus, $\Omega_\pm = 1/R$, and, in terms of the variables $(y,j)$, we have
\begin{equation}
h(y,j) = \delta(y) + \delta(y + 2j/R) .
\label{h1''}
\end{equation}
Again, for $J > 0$ the states with $y = 0$ dominate the low
temperature behavior. Since $H(0,j) = j$, eqs.(\ref{J4'}) and
(\ref{S4'}) yield
\begin{eqnarray}
J & = & \frac{1}{\sigma} \sum_{n=1}^\infty \frac{1}{n} 
\int_{0}^\infty dw  \frac{w}{n\sigma} (w-1) e^{- w}   \nonumber\\
& = & \frac{1}{\sigma^2} \zeta(2) \int_{0}^\infty dw w(w-1) e^{- w} \nonumber\\
& = & \frac{\pi^2}{6 \sigma^2} ,
\label{J8'} 
\end{eqnarray}
and, similarly,
\begin{equation}
S = \frac{\pi^2}{3 \sigma} .
\label{S8'}
\end{equation}
Thus, we find that at $T = 0$,
\begin{equation}
S = \frac{2 \pi}{\sqrt{6}} J^{1/2} ,
\label{S9'}
\end{equation}
in violation of the Nernst formulation of the third law. Note that
this example is essentially the same system as considered in the
string theory models of charged black holes which saturate the BPS
bound \cite{sv}.

Encouraged by the ability to violate the Nernst formulation of the
third law in the simple examples above, we may ask whether it is
possible to reproduce the relations (\ref{M0}) and (\ref{S0}) with an
ideal boson gas at absolute zero temperature. However, it is easy to
see that if $S(T,J)$ remains finite as $T \rightarrow 0$, then
eq.(\ref{M0}) {\em cannot} be satisfied by any ideal boson gas at $T =
0$. Namely, it follows immediately from eqs.(\ref{E}) and (\ref{S})
that as $T \rightarrow 0$, we have $(E - \Omega J) \rightarrow
0$. However, since $\sigma = \beta(\Omega_+ - \Omega)$ always remains
bounded as $T \rightarrow 0$ at fixed $J > 0$ (see eq.(\ref{sigma0})
above), we also have $\Omega \rightarrow \Omega_+$ as $T \rightarrow
0$. Thus, provided only that $S$ is finite at $T = 0$, the relation
\begin{equation}
E = \Omega_+ J
\label{E0}
\end{equation}
always holds at $T = 0$, rather than $E \propto J^{1/2}$ as in eq.(\ref{M0}).

However, a simple and natural modification of the model of a boson gas
confined to a ring does yield the desired behavior $E \propto
J^{1/2}$. Suppose that we take the ring radius, $R$, to be an
additional dynamical degree of freedom of the system (which we treat
classically). In addition, suppose that, due to tension, this ring has
an energy $E_R = \lambda R$ with $\lambda$ a constant. In other words,
suppose that the ``ring'' is actually a ``string''. (The ``massless
boson gas confined to the ring'' could then arise naturally as certain
(quantized) degrees of freedom describing deviations of the string
from circularity.) The total energy of the system would then be
\begin{equation}
E = E_G + E_R = E_G + \lambda R
\label{Etot}
\end{equation}
where $E_G$ denotes the energy of the boson gas. By eq.(\ref{E0}), at
$T = 0$ we have $E_G = \Omega_+ J = J/R$, and $R$ will be determined
by minimizing the total energy. We obtain
\begin{equation}
R = \sqrt{J/\lambda}
\label{R}
\end{equation}
and, thus
\begin{equation}
E = 2 \sqrt{\lambda} J^{1/2} ,
\label{E0'}
\end{equation}
in agreement with the behavior in eq.(\ref{M0}).

Can eq.(\ref{S0}) also be satisfied in this model? As calculated
above, for a free, massless boson gas (or a collection of such gases),
we have $S \propto J^{1/2}$ (see eq.(\ref{S9'})), rather than $S
\propto J$, as required by eq.(\ref{S0}). Indeed, for any system for
which eqs.(\ref{J4'}) and (\ref{S4'}) hold at $T = 0$ and for any
polynomial behavior of $H(0,j)$ such that $H(0,0) = 0$ (see
eq.(\ref{yj'})), it is easy to check that $S/J \rightarrow 0$ as $J
\rightarrow \infty$. What seems to be required to obtain the behavior
(\ref{S0}) in any model where eqs.(\ref{J4'}) and (\ref{S4'}) hold at
$T = 0$ is to have exponential growth of $H(0,j)$ at large $j$. I know
of no physically reasonable model involving an ideal boson gas in
which this behavior occurs.

Nevertheless, one possibility is worth analyzing further with regard to
whether the behavior (\ref{S0}) at $T = 0$ can be obtained in the
above simple ``string model''. Suppose we allow the string to have a
spectrum of massive particles which rises exponentially in $M$, i.e.,
$n(M) \propto e^{\alpha M}$. (Such an exponentially rising spectrum
actually occurs in string theory.) Although for a massive particle, no
single particle states satisfy $j/\epsilon = R$, a sufficiently
rapidly growing density of states -- in particular, as discussed
above, exponential growth of the density of states in $j$ at fixed $y$
-- could allow states with $j/\epsilon < R$ to contribute to the
thermodynamic properties of the system at $T = 0$, thus invalidating
eqs.(\ref{J4'}) and (\ref{S4'}). Since each particle of mass $M$
contributes a density of states $g_M(\epsilon,j) = \delta(\epsilon -
\sqrt{M^2 + j^2/R^2})$, the density of states for an exponentially
rising spectrum behaves as
\begin{equation}
g(\epsilon,j) \sim e^{\alpha \sqrt{\epsilon^2 - j^2/R^2}}
\label{gM}
\end{equation}
or, equivalently,
\begin{equation}
h(y,j) \sim e^{\alpha \sqrt{y^2 + 2yj/R}}
\label{hM}
\end{equation}
The leading order behavior of $H(y,j)$ (at large values of $y^2 +
2yj/R$) is similar. Thus, $H(y,j)$ does indeed grow more rapidly than
polynomially in $j$ at fixed $y$, but it also grows more slowly than
exponentially in $j$. If any massless particles are present (so that
$H(0,j) > 0$ for some $j$), then $\sigma_0 > 0$, and it is not
difficult to see that the massive states will not, in fact, contribute
to the thermodynamic behavior of the system at $T = 0$. On the other
hand, if no massless particles are present, then the growth of states
with $j$ is not rapid enough to avoid having $\sigma_0 = 0$, and the
Nernst formulation of the third law should hold. Thus, I see no
natural way of obtaining the behavior (\ref{S0}) at $T = 0$ in the
context of this simple ``string model''. Of course, as emphasized in
the Introduction, we have little right to expect to be able to obtain
all of the thermodynamic properties of extremal rotating black holes
with such a naive model.

This research was supported in part by NSF grant PHY 95-14726 to the
University of Chicago.

\end{document}